\documentclass[aip,reprint,cha]{revtex4-1}

\usepackage[utf8]{inputenc}
\usepackage[T1]{fontenc}

\newcommand{\layDense}{\mathop{\mathrm{Dense}}}
\newcommand{\layInput}{\mathop{\mathrm{Input}}}
\newcommand{\layTake}{\mathop{\mathrm{Take}}}
\newcommand{\laySans}{\mathop{\mathrm{Sans}}}
\newcommand{\layScale}{\mathop{\mathrm{Scale}}}
\newcommand{\sansU}[1]{u_{\neg\,#1}}
\newcommand{\euler}{\mathrm{e}}

\usepackage{graphicx}
\usepackage{amsmath}
\usepackage{amssymb}

\begin{document}


\title{Reconstruction of neuromorphic dynamics from a single scalar time series
  using variational autoencoder and neural network map} 

=


\author{Pavel V. Kuptsov}
\email{kupav@mail.ru, corresponding author}
\altaffiliation[also at ]{Kotelnikov Institute of Radio-Engineering and
  Electronics of Russian Academy of Sciences, Saratov Branch, 38 Zelenaya str.,
  Saratov 410019, Russia}%
\author{Nataliya V. Stankevich}
\email{stankevichnv@mail.ru}
\altaffiliation[also at ]{Kotelnikov Institute of Radio-Engineering and
  Electronics of Russian Academy of Sciences, Saratov Branch, 38 Zelenaya str.,
  Saratov 410019, Russia}%
\affiliation{Laboratory of topological methods in dynamics, HSE University,
  25/12 Bolshaya Pecherskaya str., Nizhny Novgorod 603155, Russia}


\date{\today}

\begin{abstract}
  This paper examines the reconstruction of a family of dynamical systems with
  neuromorphic behavior using a single scalar time series. A model of a
  physiological neuron based on the Hodgkin-Huxley formalism is
  considered. Single time series of one of its variables is shown to be enough
  to train a neural network that can operate as a discrete time dynamical system
  with one control parameter. The neural network system is created in two
  steps. First, the delay-coordinate embedding vectors are constructed form the
  original time series and their dimension is reduced with by means of a
  variational autoencoder to obtain the recovered state-space vectors. It is
  shown that an appropriate reduced dimension can be determined by analyzing the
  autoencoder training process. Second, pairs of the recovered state-space
  vectors at consecutive time steps supplied with a constant value playing the
  role of a control parameter are used to train another neural network to make
  it operate as a recurrent map. The regimes of thus created neural network
  system observed when its control parameter is varied are in very good
  accordance with those of the original system, though they were not explicitly
  presented during training.
\end{abstract}


\maketitle 

\begin{quotation}
  One of the most appealing characteristics of neural networks is their ability
  for data generalization. This is achieved through the extraction of data
  features and dependencies that may initially appear obscure. In the context of
  dynamical systems reconstruction, this aptitude could facilitate significant
  advancements in the development of models based on experimental data. In this
  paper, we investigate the extent to which neural networks can reproduce the
  dynamical regimes of a system, observed for various values of its control
  parameters, when only a single scalar time series of this system is available.
  Created neural network models a family of dynamical systems parameterized by a
  control parameter. This family exhibits behavior consistent with that of the
  original system, i.e. discovers its specific regimes and transitions. This is
  demonstrated by the example of a neuromorphic Hodgkin-Huxley system. The
  reconstruction is carried out in two steps. Initially, a variational
  autoencoder is trained to convert the time series into a sequence of
  reconstructed state vectors. These are then used for training a neural network
  map. The trained map operates as a dynamical system with discrete time, having
  one control parameter. We show that it is capable of reproducing the dynamical
  regimes of the original system, including those that were not demonstrated
  during training.
\end{quotation}

\section{Introduction}

Reconstructing dynamical models from time series is an important task from a
practical point of view and also is an interesting theoretical challenge. To
date, a large number of approaches and methods for solving this problem have
been developed~\cite{kantz2004,Kantz2015,BezSmir2010}. In connection with the
intensive development of artificial neural network technologies, it seems
interesting to study their capabilities in relation to the dynamical
modeling. In this regard, without claiming completeness, we can mention some
related works. In papers~\cite{Shilnikov2021,Kong2021,Hart2024} a sort of neural
networks techniques called reservoir computing is considered as tool for
replicating the dynamics of chaotic systems and predicting their
properties. Blending several dynamical systems into a single model built on the
bases of reservoir computing is considered in
papers~\cite{Flynn2021,Flynn2023}. In paper~\cite{Durstewitz2023} recurrent
neural network is considered as a tool for reconstruction of a model of dynamics
trained directly on the measured physiological data. Paper~\cite{Zhang2019}
reports a new framework, Gumbel Graph Network, which is a model-free,
data-driven deep learning framework to accomplish the reconstruction of both
network connections and the dynamics on it. Paper~\cite{Almazova2021} is devoted
to reconstruction of chaotic dynamic from scalar time series using plain
autoencoder.

Along with the reconstruction of dynamical systems, neural networks and
machine learning methods inspire new ideas in other fields of nonlinear
dynamics. In particular, a series of
papers~\cite{NewSyncDynLearn,NewSyncFindSync,NewSyncDynMod} develops a technique
of dynamical learning of synchronization (DLS) designed to control the
synchronization state of a complex network composed of nonlinear
oscillators. Paper~\cite{NewSyncEnhan} demonstrates that a three-layer neural
network, comprising input, training, and output layers and utilizing
spike-timing-dependent plasticity (STDP), can effectively learn the orderly
propagation of signals between network layers under controlled synaptic
conditions. Neural networks are used for adaptive modeling and control of
dynamical systems~\cite{Levin1991185}, applied for {E}l {N}i\~{n}o attractor
reconstruction~\cite{GrigLatif94}, employed for dynamics prediction based on
state space reconstruction~\cite{BolNak2000}. Also machine learning is utilized for
biological models creations and analysis~\cite{GILPIN20201}.

One of the most appealing characteristics of neural networks is their capacity
for data generalization. This ability stems from the fact that a well-trained
neural network is able to identify and extract from the data features and
dependencies that may initially appear to be obscure. In the context of problems
related to dynamics reconstruction, this aptitude should facilitate significant
advancements in the direction of developing models based on experimental
data. It is natural to expect that even with a limited amount of information
about the system, such as its single scalar time series, it will be possible to
use a neural network to construct a family of models, parameterized by control
parameters, that will be able to predict a variety of system behaviors.

Our paper considers this possibility in relation to a specific example: the
neuromorphic dynamics of a system built on the basis of the Hodgkin-Huxley
formalism. We avoid using simplified neuron models and focus on the realistic
Hodgkin-Huxley one since planing in our future works to consider an experimental
data obtained from a natural system described within the Hodgkin-Huxley
framework.

The main motivation of this paper is to contribute to the methods of studying
dynamics based on experimental data. However, in this paper we define a modeled
system using ODEs and get a time series from numerical solving the equations. It
means that our data are devoid of the features inherent in the data of a real
experiment: they are sampled with a small and constant time step, there are
quite a lot of them, the readings are recorded without errors (except for
numerical ones) and they are not noisy. The cases with non-perfect experimental
data will be considered in our subsequent studies.

We use variational autoencoder (VAE)~\cite{Kingma2019,Chollet2021} and neural
network map~\cite{NNDiscov22,NNNCPL2024} as tools to perform reconstruction. The
VAE is used to recover system state vectors by reducing the dimension of
delay-coordinate embedding vectors from the original time series. The recovered
state-space vectors are then used to train the neural network map, resulting in
a reconstructed dynamical system with a single control parameter.

An autoencoder is one of the standard neural network architectures, which is
used, among other things, to reduce the dimension of
data~\cite{Chollet2021}. The main feature of an autoencoder is that the data
vectors it process at some point passes through a bottleneck, that is, it turns
out to be represented in the form of low-dimensional vectors. The first part of
the autoencoder, called the encoder, is responsible for this. After this, the
dimension is restored to the original value by the second part, called the
decoder. The objective of training is to achieve a minimal discrepancy between
the input and output vectors. To this end, all pertinent data must be compressed
into the reduced vectors at the bottleneck.

VAE differs from a plain autoencoder in that the output of the encoder is not
the reduced vector itself, but parameters of the so-called latent space. The
reduced vector is randomly selected from this latent space and passed to the
decoder. Random sampling of the reduced vectors from the latent space during
training is very important for the purposes of dynamics reconstruction, as it
ensures proximity preservation. Thanks to this, it is guaranteed that fragments
of the original trajectory that are close to each other will be represented by
the close latent space vectors, which we use as the reconstructed state-space
vectors of the modeled system.

It is mathematically proven~\cite{Kolmogorov56, Arnold57, Kolmogorov57,
  Cybenko1989, Gorban1998, Haykin1998} that a two-layer fully connected, i.e.,
dense, neural network can be used to approximate arbitrary functions of many
variables. It means that it can be suitable for reconstruction of a dynamical
system. In paper~\cite{NNDyn21} a good quality of dynamics reproduction with
such a simple network model was shown for different systems: Lorenz,
R\"{o}ssler, Hindmarch–Rose neuron model. However, already for the
Hindmarch–Rose oscillator the quality of reconstruction was lower that for two
other systems. It was found that a neural network map based on a simple
two-layer structure is difficult to train for so called stiff systems where
variables have very different time scales. Then in paper~\cite{NNDiscov22} a
more advanced structure of the neural network map was suggested. Instead of a
single network accepting the full state vector at once and returning the state
vector on the next time step, subnetworks for each dynamical variable were
introduced. Each subnetwork still has two-layer structure, but all other
variables together with control parameters are injected after an additional
dense layer. Such network is found to model stiff dynamics and also is able to
discover correctly dynamical regime not shown during training. In
paper~\cite{NNNCPL2024} it is shown that two neural network maps trained
separately can model dynamics of two coupled systems without additional
training.

Thus the goal of this study is to reconstruct the dynamics of a system as
completely as possible, given its single scalar time series, using neural
networks. The result of the reconstruction will be a family of dynamical
systems, parameterized by a control parameter. Our particular interest will be
in the dynamical features of the modeled system that are not explicitly
presented during training and nevertheless discovered by the reconstructed
system.

\section{Hodgkin-Huxley-type model}

We consider a model of a physiological neuron based on the Hodgkin-Huxley
formalism~\cite{Sherman1988}, and also modification~\cite{StanMos2017} that is
responsible for bistability in a certain parameter range when bursting
oscillations coexist with a stable equilibrium point,
\begin{equation}\label{eq:sys}
  \begin{aligned}
    \tau \dot V &= -I_{Ca}(V) - I_{K}(V,n)-I_S(V,S)-kI_{K2}(V),\\
    \tau \dot n &= \sigma\,[n_{\infty}(V)-n],\\
    \tau_{S} \dot S &= S_{\infty}(V)-S.
  \end{aligned}
\end{equation}
Here $V$, $n$ and $S$ are dynamical variables. In the considered parameter
ranges the system~\eqref{eq:sys} has a single unstable fixed point. The
modification resulting in its stabilization is introduced via the function
$I_{K2}(V)$. It will be switched on and off by the coefficient $k\in\{0,1\}$. We
will consider the system~\eqref{eq:sys} for varying $V_S$ at $k=0$ (the original
system) and $k=1$ (the modified system). All other numerical values of
parameters as well as the functions included in these equations are given in
Appendix~\ref{sec:sys_detail}, see Table~\ref{tab:param} and
Eqs.~\eqref{eq:origsys_funcs}.

System~\eqref{eq:sys} is stiff since it has one slow and two fast variables with
very different rates of oscillations. Effective numerical solution of a system
like this is possible with specially designed methods. We will use a stiff
solver implementing Radau method~\cite{Hairer1999}.

\begin{figure}
  \centering\includegraphics{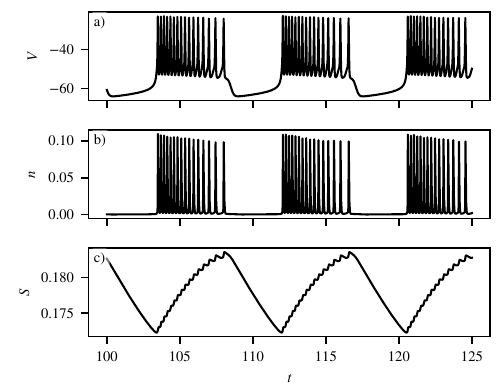}
  \caption{\label{fig:sol_sherman_ode_o_burst}Solution of the
    system~\eqref{eq:sys} in the regime of bursting oscillations. $k=0$ and
    $V_S=-36$. Other parameters see in Tab.~\ref{tab:param}.}
\end{figure}

\begin{figure}
  \centering\includegraphics{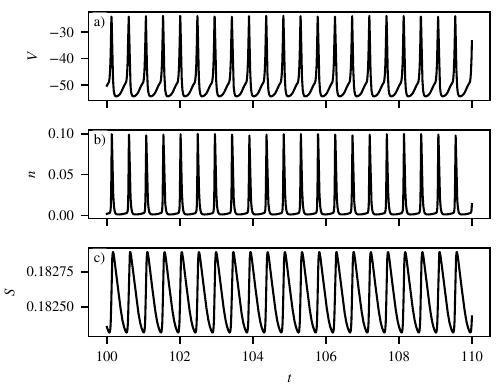}
  \caption{\label{fig:sol_sherman_ode_o_spikes}Spiking of the
    system~\eqref{eq:sys} at $k=0$ and $V_S=-33$; also see
    Tab.~\ref{tab:param}.}
\end{figure}

Dynamics of the system~\eqref{eq:sys} at $k=0$ is illustrated in
Figs.~\ref{fig:sol_sherman_ode_o_burst}
and~\ref{fig:sol_sherman_ode_o_spikes}. Figure~\ref{fig:sol_sherman_ode_o_burst}
represents busting oscillations at $V_s=-36$. Observe that $V$ and $n$ oscillate
fast while $S$ is a slow variable. When $V_S$ gets larger spiking regime emerges
as shown in Fig.~\ref{fig:sol_sherman_ode_o_spikes} that is plotted at
$V_S=-33$. Solutions of the modified system at $k=1$ are not shown since they
are visually indistinguishable from those in
Figs.~\ref{fig:sol_sherman_ode_o_burst} and \ref{fig:sol_sherman_ode_o_spikes}.

\begin{figure}
  \centering\includegraphics{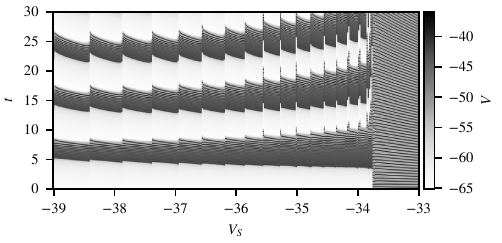}
  \caption{\label{fig:bif_diag_sherman_ode_o}Diagram of regimes of the
    system~\eqref{eq:sys} at $k=0$. The horizontal axis shows the change in
    parameter $V_S$. For each $V_S$ a solution of the system~\eqref{eq:sys} is
    computed and values of $V(t)$ are shown along the vertical axis with gray
    shapes. The darker shades indicate higher values. The plotted solutions are
    recorded after omitting transients and are aligned by the smallest minimum
    on the curve $V(t)$.}
\end{figure}

Figure~\ref{fig:bif_diag_sherman_ode_o} shows how dynamics of the
system~\eqref{eq:sys} at $k=0$ changes. Here $V_S$ varies along the horizontal
axis, time goes vertically and shades of gray indicate values of dynamical
variable $V$ recorded after omitting transients and aligned by the smallest
minimum on the curve $V$. Burst in the left part of the figure looks like three
stripes that vanish at $V_S\approx -33.73$ due to the bifurcation transition
associated with a blue-sky catastrophe~\cite{StanMos2017,ShilCymb2005}. To the
right of this point high frequency spiking oscillations appear that are
represented in the diagram as a uniform texture.

\begin{figure}
  \centering\includegraphics{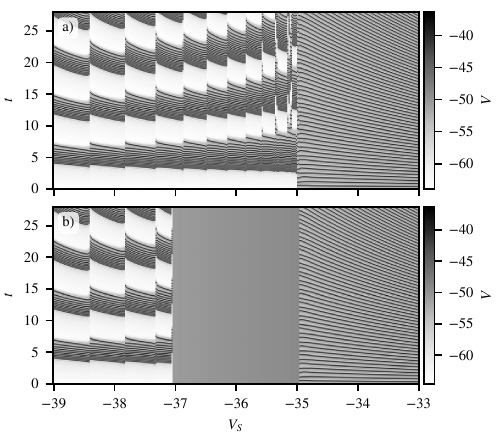}
  \caption{\label{fig:bif_diag_sherman_ode_m}Diagram of regimes of the
    system~\eqref{eq:sys} at $k=1$. Panels (a) and (b) illustrate the
    bistability. To plot each panel we compute one of two coexisting solutions
    in the middle of the plot at $V_S=-36$: (a) the bursts and (b) the fixed
    point. Then this solution is continued to the left and to the right, i.e.,
    computation at new parameter step starts from the last point obtained at the
    previous step}
\end{figure}

Bistability at $k=1$ is shown in Fig.~\ref{fig:bif_diag_sherman_ode_m}. For
$V_S$ approximately between $V_S=-37$ and $V_S=-35$ there is a stable fixed
point. At $V_S=-36$ the fixed point is $V=-50.636$, $n=2.0560\times 10^{-3}$,
$S=0.18792$, and its stability is indicated by the negative eigenvalues
($-0.15927$, $-19.521$, $-38.785$). The stable fixed point coexists with the
bursting oscillations. To draw Fig.~\ref{fig:bif_diag_sherman_ode_m} we begin at
$V_S=-36$ and continue the solution step by step first to the left and to then
the right of this parameter value always taking the last trajectory point at the
previous parameter step as the initial point for the next parameter step.  In
Fig.~\ref{fig:bif_diag_sherman_ode_m}(a) the very starting point for the
trajectory at $V_S=-36$ arrives after a transient at the bursting
oscillations. Doing continuation of this solution we observe a picture that is
very similar to the system at $k=0$,
cf. Fig.~\ref{fig:bif_diag_sherman_ode_o}. And when the very starting point is
taken near the stable fixed point, as in
Fig.~\ref{fig:bif_diag_sherman_ode_m}(b), the flat area appears in the figure
that represents the staying the system at the fixed point.

\section{\label{sec:VAE}Reconstruction state-space vectors using variational
  autoencoder}

\subsection{Approach in general}

Autoencoders are neural networks that are trained via self supervised
procedure~\cite{Kingma2019,Chollet2021}. The training vectors $R$ are forwarded
to the network and the goal of the training is to restore at the output the
vectors $R'$ as close as possible to $R$. The loss function of the training is
the distance between $R$ and $R'$. The key point is the presence of a
bottleneck. The autoencoder consists of two parts, an encoder and a decoder. The
encoder maps the input vectors $R$ to the space of reduced vectors $u$ of
smaller dimension, and the decoder maps them back to the original space of
$R$. Due to the squeezing the data through the low dimensional space, the
network is forced to extract the most essential data features sufficient for
their further restoring.

In plain autoencoders the encoder output $u$ goes directly to the decoder
input. VAE encoder generates parameters of so called latent space, usually they
are a mean value $\mu$ and a standard deviation $\sigma$ of a Gaussian
distribution. The decoder input vector $u$ is a random sample from this
distribution. The mathematics theory of VAEs can be found in
paper~\cite{Kingma2019} and for the software implementation we have followed the
description form the book~\cite{Chollet2021}.

We will use VAEs for state-space vector reconstruction. At first glance, random
sampling from the latent space introduces unnecessary stochasticity to
reconstruction procedure. Plain autoencoders that do not use the random sampling
can also be used for dimension reduction of phase space
vectors~\cite{Kuptsov2019,Almazova2021}. However, plain autoencoders do not
guarantee a continuity of the space of the reduced vectors. Close input vector
will not necessarily be mapped to the close vectors in the latent space merely
since no limitation is applied to the training process that would force this
property to appear. For the dynamic reconstruction it means that given close
parts of the modeled trajectory there are no reasons to expect that the encoder
will always map them to close reconstructed state-space vectors. The proximity
preservation in the latent space is very important since without it the
reconstructed system will be inappropriate for method of stability analysis
based on small perturbations, like computation of Lyapunov exponents. It is the
random sampling from the latent space that provides the very training constraint
that ensures the proximity preservation~\cite{Kingma2019}. Thanks to it, not
only the exact vector $u=\mu$ will be mapped back to $R'\approx R$ but also its
small perturbations will also be mapped to vectors close to $R$. Proximity
preservation can be treated as a sort of robustness: two similar system will
obtain similar models, and a small neighborhood of the original trajectory will
be mapped to a small neighborhood of the trajectory of the reconstructed system.

Consider a dynamical system given by ODEs:
\begin{equation}
  \dot U = f(U),
\end{equation}
where $U=U(t)$ is a state-space vector of dimension $D_u$. This equation is
solved numerically to obtain a discrete time series of vectors with constant
time step $\Delta t$. Then only one scalar component
$U_i(n)=U_i(t_0+n\Delta t)$, $i\in\{1,2,\ldots, D_u\}$, is taken for further
processing. The choice of $i$ is a subject of a separate discussion.

This step models an experimental situation when the full vector as well as an
underlying dynamical system are unknown and the only scalar time series is
available. In this paper we do not consider a realistic situation when the data
record is written with noises, with large discretization step and is too
short. We just take ``good'' numerical data and left more realistic cases for
further studies.

The standard first step of data preparation is the rescaling of $U_i(n)$ to fit
the range $[-1,1]$ for the neural network training algorithm to work
properly. Then this data series is transformed to delay-coordinate embedding
vectors~\cite{Packard1980, Takens1981} $R(n)$ of dimension $D_e$:
\begin{equation}
  \label{eq:embed_vector}
  R(n) = \big(U_i(n), U_i(n-1), \ldots , U_i(n-D_e+1)\big).
\end{equation}
The dimension $D_e$ is usually unknown in model reconstruction procedure. There
are rather heuristic algorithms of their
determinations~\cite{kantz2004,Takens1981}. We will not use them since dimension
reduction procedure via VAE is not sensible to the very exact choice of
$D_e$. It is enough to take it excessively large.

Vectors $R$ are used to train the VAE whose detailed structure is described in
Appendix \ref{sec:vae_detail}. The result is a two-step map. The first part, the
encoder, maps $R$ to two vectors $\mu$ and $\sigma$ of the reduced dimension
$D_u$. The second part, the decoder, maps a random sample $u$ of a Gaussian
distribution $N(\mu,\sigma)$ to a vector $R'$ that is approximately equals to
$R$.

\subsection{\label{sec:vae_hodg_hux}Variational autoencoder for the
  Hodgkin-Huxley-type model}

We are going to model the periodic bursting oscillations of the
system~\eqref{eq:sys}, like those shown in
Fig.~\ref{fig:sol_sherman_ode_o_burst}. The information about the modeled system
is limited by a single period. The period size depends on the parameters and
approximately equals to $T\approx 10$. Step size of time discretization is
$\Delta t = 0.005$. The scalar time series of one period length is sliced into
the delay-coordinate embedding vectors $R(n)$ of dimension $D_e=32$, see
Eq.~\eqref{eq:embed_vector}.

The approach we develop is not sensible to the particular choice of $D_e$
provided that it excessively larger then dimension of the considered system
$D_u=3$. This is the advantage of using VAE for state vector recovery. However
to make sure we checked various $D_e$: $16$, $32$, $64$, $128$ and $512$, and
all worked equally well.

The approximate size of the resulting dataset is $T/\Delta t\approx 2000$ which
is small. Training a neural network on a small dataset usually results in the
network simply memorising the data instead of the desired generalization of the
data, i.e. the extraction of their essential features. One of the standard
methods to improve this situation is data augmentation. We perform it as
follows: the trajectory cut of one period length is supplied with its $19$
copies perturbed by a small Gaussian noise with zero mean and the standard
deviation $1/200$. Besides the improved data generalization the augmentation
serves for better robustness: the obtained VAE is trained to encode and decode
not the unique trajectory, but a small cloud of them.

\begin{figure}
  \centering\includegraphics{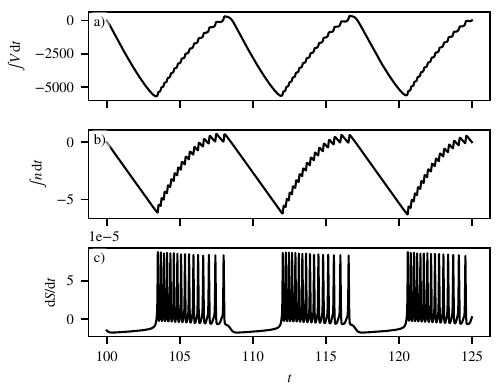}
  \caption{\label{fig:sol_sherman_ode_o_burst_integ}Integrated $V(t)$ and
    $n(t)$, panels (a) and (b), respectively, and differentiated $S(t)$, panel
    (c), taken form Fig.~\ref{fig:sol_sherman_ode_o_burst}.}
\end{figure}

The result of the VAE training is found to depend on the choice of the scalar
variable for reconstruction. Figure~\ref{fig:sol_sherman_ode_o_burst_integ}
shows that integration of the fast variables $V$ and $n$ of the
system~\eqref{eq:sys} results in slow curves that look very similar to $S$. The
differentiating of $S$ produces the bursting curve whose shape is very similar
to $V$. As already discussed above a scalar time series is converted to
delay-coordinate embedding vectors $R(n)$ of the dimension $D_e$. The network
processes each of these vectors separately so that it can operate only with
trajectory pieces of the length $D_e$. It means that VAE cannot integrate the
whole input time series to find a slow variable given a fast one. But it can
differentiate data since this is a local operation so that all necessary
information is available for it. Moreover, the convolution layer employed in the
encoder, see Fig.~\ref{fig:vae_arch}(b), can perform the differentiation
provided that its weights are tuned properly.

\begin{figure*}
  \centering\includegraphics{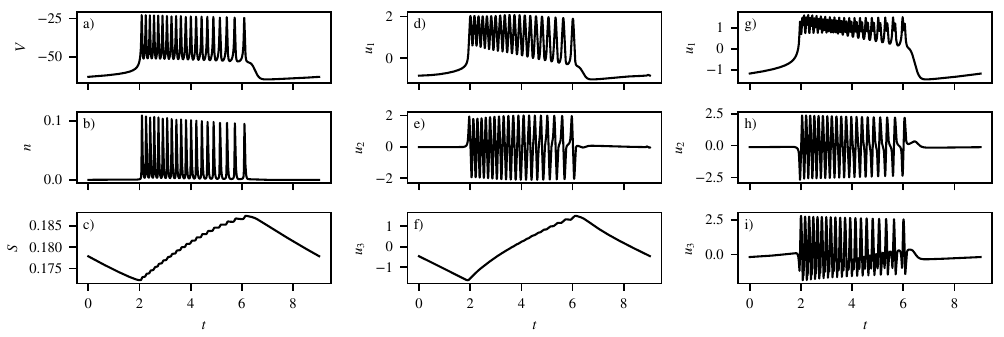}
  \caption{\label{fig:sol_sherman_vae_sv}Panels (a), (b) and (c) depict one
    period of a bursting trajectory of the system~\eqref{eq:sys} at $V_S=-36$
    and $k=0$. Panels (d), (e) and (f) illustrate output of the VAE encoder
    trained on the time series for $S(t)$ from the panel (c). Panel (g), (h) and
    (i) --- output of the VAE encoder trained on $V(t)$ from the panel (a).}
\end{figure*}

As a result when $S$ is used for training VAE, its encoder is able to
reconstruct two fast and one slow variable and being trained for $V$ or $n$ it
cannot reconstruct a slow variable. This is shown in
Fig.~\ref{fig:sol_sherman_vae_sv}. Panels (a-c) demonstrate initial data
obtained as a solution of Eqs.~\eqref{eq:sys} at $V_s=-36$ and $k=0$ on a single
period $T=9$ with the time step $\Delta t = 0.005$. Panels (d-f) show vectors
$u(n)=\mu(n)$ obtained at the VAE encoder output that is forwarded by the
embedding vectors obtained for $S$. Observe a very good correspondence of $V$
and $u_1$ and also $S$ and $u_3$. As for $n$ its reconstructed version $u_2$
demonstrates similar burst and the only difference is that $u_2$ oscillates up
and down with respect to the middle level while for $n$ there is a minimum level
and bursts occur only above it. Nevertheless VAE encoder performance is good as
a whole: panels (a-c) and their reconstruction are qualitatively similar to each
other. Notice that the variables $u_1$, $u_2$ and $u_3$ appears in an arbitrary
order. We reorder them manually to fit the sequence $V$, $n$ and $S$ of the
system~\eqref{eq:netw_model}. Quite different result is obtained for VAE trained
on $V$, see panels (g-i). Variables $V$ and $n$ correspond to the reconstructed
variables $u_1$ and $u_2$, but no slow variable similar to $S$ appears. Instead
$u_3$ demonstrate bursts rather similar to $u_2$. Thus in what follows we will
take time series of $S(n)$ to create a reconstructed system.

When a dynamical system is represented by its scalar time series the true value
of the dimension $D_u$ is unavailable. When performing state vector
reconstruction using a plain autoencoder, one way to determine it is to impose
an additional constraint on the vectors in the latent space by requiring that
their sparsity to be maximised during training. This method is used in
paper~\cite{Almazova2021}. We do not employ this approach since we consider a
variational autoencoder. Its encoding part has two outputs $\mu$ and $\nu$ that
are already constrained by the minimization of the Kullback–Leibler divergence,
see Eq.~\eqref{eq:vae_loss} and the paper~\cite{Kingma2019} for corresponding
theoretical considerations. Adding more constrains to $\mu$ and $\nu$ would
require a modification of the mathematical foundation of variational
autoencoders. Due to this reason we will merely build VAEs for different latent
space dimension $D_u$ and discuss how the guess of $D_u$ influences the result
of the reconstruction.

\begin{figure*}
  \centering\includegraphics{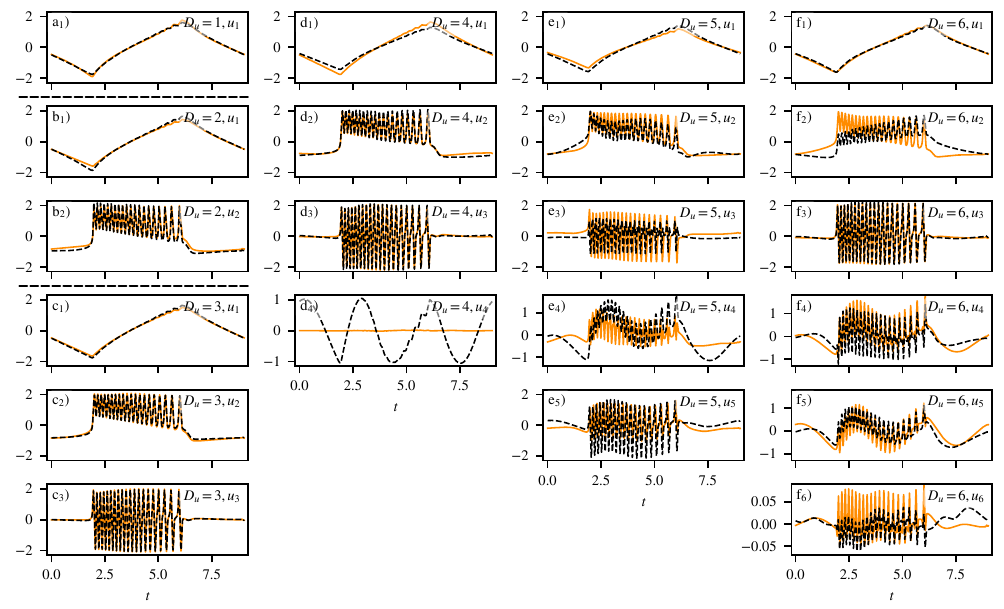}
  \caption{\label{fig:vae_vari_du}Time series of the reconstructed state-space
    vectors $u_i$ ($i=1,2,\ldots, D_u$), i.e., VAE encoder outputs, obtained when VAE
    is trained on the same time series $S(t)$ at various $D_u$:
    (a$_1$) $D_u=1$; %
    (b$_1$) and (b$_2$) $D_u=2$; %
    (c$_1$)--(c$_3$) $D_u=3$; %
    (d$_1$)--(d$_4$) $D_u=4$; %
    (e$_1$)--(e$_5$) $D_u=5$; %
    (f$_1$)--(f$_6$) $D_u=6$. %
    The dashed horizontal lines between axis separate panels in the left column
    with different values of $D_u$.
    The training time series $S(t)$ is shown in Fig.~\ref{fig:sol_sherman_vae_sv}(c),
    Solid orange and dashed black curves on each panel illustrate two runs of
    VAE training with identical parameters and random initial weights. The
    curves for each VAE have been manually reordered to achieve correspondence
    between the curves for different $D_u$.}
\end{figure*}

Figure~\ref{fig:vae_vari_du} shows VAE encoder
outputs computed for the same input data: one period of the variable $S$ of the
system~\eqref{eq:sys} at $V_S=-36$ and $k=0$, see
Fig.~\ref{fig:sol_sherman_vae_sv}(c). VAEs are trained for six guesses of the
dimension $D_u=1,2,\ldots 6$, moreover, for each $D_u$ the training is performed
two times starting from random initial weights $w_e$ and $w_d$ of the network.
Totally the results of twelve VAEs are shown.

First of all notice that the components $S$ that is used as training data is
reconstructed correctly for all guesses of $D_u$ and for all runs of training,
see Figs.~\ref{fig:vae_vari_du}(a$_1$,b$_1$,\ldots, f$_1$).

When the dimension guess is enough for the second component to appear,
$D_u\ge 2$, VAE encoders start to reconstruct the variable $V$, see
Figs.~\ref{fig:vae_vari_du}(b$_2$,c$_2$,\ldots, f$_2$). Similarly when
$D_u\ge 3$ the component corresponding to $n$ is recovered,
Figs.~\ref{fig:vae_vari_du}(c$_3$,d$_3$\ldots, f$_3$). Notice that when
$D_u\le 4$ the recovering of ``true'' components of the system~\eqref{eq:sys}
$S$, $V$ and $n$ is stable: curves for different runs of training are almost
identical, compare solid and dashed curves in Figs.\ref{fig:vae_vari_du}(a$_1$,
b$_{1,2}$, c$_{1,2,3}$ and d$_{1,2,3}$). When $D_u= 5$ or $6$, the different
runs of the training results in not completely identical curves for $V$ and
$n$. Nevertheless they are still very similar, compare solid and dashed curves
in Figs.\ref{fig:vae_vari_du}(e$_{2,3}$ and f$_2$). Summarizing, VAE firmly
reproduces ``true'' components of the modeled system regardless of dimension
$D_u$ guess. However when $D_u$ gets higher the reproduction accuracy
deteriorates. When the guess for $D_u$ is higher then the true one, spurious
components appear. Analyzing the corresponding curves in
Figs.~\ref{fig:vae_vari_du}(d$_4$), ~\ref{fig:vae_vari_du}(e$_{4,5}$) and
\ref{fig:vae_vari_du}(f$_{4,5,6}$) one sees that their shapes are unstable,
i.e., they are different for different training runs, they are not reproduced
with increasing $D_u$ and one cannot find a pattern of their emerging.

\begin{figure}
  \centering\includegraphics{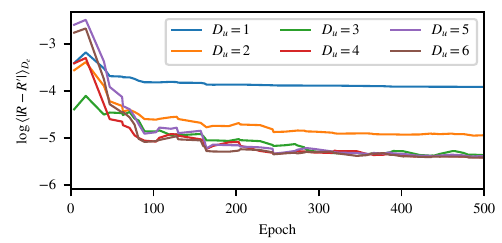}
  \caption{\label{fig:vae_learn_curves}Logarithms of mean absolute
    reconstruction error $\langle |R-R'|\rangle_{D_e}$ as a function of epoch of
    training for six VAEs trained for various $D_u$.}
\end{figure}

Influence of the guess for $D_u$ on VAEs performance is show in
Fig.~\ref{fig:vae_learn_curves} that demonstrates dependence of mean absolute
reconstruction error (MAE) $\langle |R-R'|\rangle_{D_e}$ on the epoch of
training. When the guess grow, $D_u=1$, $2$, $3$, the error becomes smaller. One
clearly see that $D_u=1$ and $2$ are not enough for correct model reconstruction
due to high error. But when $D_u$ goes beyond the true value $D_u=3$ no further
decrease of the error observed. Thus the increase of the latent space dimension
improves VAE performance only until the dimension is less then the true one.

Therefore, the observations represented in Figs.~\ref{fig:vae_vari_du}
and~\ref{fig:vae_learn_curves} demonstrate that given a scalar time series VAEs
are able to find the true dimension of an underlying dynamics.

\section{\label{sec:ntw}Neural network map}

\subsection{Structure of the neural network map}

Assume that there is a time series $u(n)$ of state-space vectors of a dynamical
system sampled at time moments $t_0+n\Delta t$, where $\Delta t$ is a step of
time discretization.  We want to create a model for this dynamical system and
for this purpose we find a function $F(\cdot)$ that maps $u(n)$ to $u(n+1)$:
\begin{equation}
  \label{eq:gen_netw_model}
  u(n+1)=F(u(n), p, w_m),
\end{equation}
Here $w_m$ is a vector of parameters that have to be tuned to adopt the
recurrent map~\eqref{eq:gen_netw_model} for the desired behavior. Vector $p$
collects control parameters of the dynamical system~\eqref{eq:gen_netw_model}
introduced to model not only the precise time series $u(n)$ but also regimes
when parameters are varied.

We will create the map~\eqref{eq:gen_netw_model} as a neural network that
operates according to the explicit formula~\cite{NNDiscov22,NNNCPL2024}:
\begin{equation}
  \label{eq:netw_model}
  \begin{split}
    u_i(n+1) &= u_i(n) + \chi \bigg(f\Big(u_i(n) a_i + \mu_i \\
    &+g\big(\sansU{i}(n) A_i + p B_i + \beta_i\big)\Big) b_i + \gamma_i\bigg).
  \end{split}
\end{equation}
This equation is build of scalars, vectors and matrices that are described in
Tab.~\ref{tab:netw_dims} in Appendix~\ref{sec:netw_map_detail}. Vector of
dynamical variables $u(n)$ has dimension $D_u$. Scalar $u_i(n)$ denotes its
$i$th component and represents a dynamical variable of the $i$th
subsystem. Vector $\sansU{i}(n)$ is obtained from $u(n)$ by removing the $i$th
component and thus has dimension $D_u-1$.  As usual for neural networks notation
$u(n)$ and $\sansU{i}(n)$ are treated as row-vectors. Control parameters are
collected as a $D_p$-dimensional row-vector $p$. Functions $f(\cdot)$ and
$g(\cdot)$ are called activation functions. They are scalar functions of scalar
arguments and when they are applied to vectors, elements-wise operation is
assumed.

Coefficient $\chi$ is a small constant. It is required to stabilize the dynamics
of the map~\eqref{eq:netw_model}. Without it the map even after a successful
training can sometimes demonstrate the divergence. This is because the neural
network map~\eqref{eq:netw_model} has a structure of the Euler scheme of
numerical ODE solving with $\chi$ playing the role of a time step. This scheme
is known to diverges if the time step is too large~\cite{Recipes}. Thus, the
search for a suitable value of $\chi$ reduces to finding the maximum time step
at which the corresponding Euler scheme for Eq.~\eqref{eq:sys} remains
stable. Straightforward check reveled that the condition $\chi<0.01$ is enough
for the stability, and thus for further computations we set
$\chi=\Delta t=0.005$.

Detailed description of the neural network map as well as discussion of its
software implementation can be found in Appendix~\ref{sec:netw_map_detail}.

\subsection{Neural network map for the reconstructed space-state vectors of the
  Hodgkin-Huxley-type model}

To prepare the training data for the neural network map~\eqref{eq:netw_model} we
again take the embedding vectors $R(n)$, see Eq.~\eqref{eq:embed_vector} and
forward them, now without the augmentation, to the VAE encoder. Its output is
used to prepare a time series of reconstructed state-space vectors $u(n)$. These
vectors are element-wise scaled to fit the range $[-1,1]$ for standard neural
network training algorithms to work properly. Sequence of the resulting vectors
correspond to time steps $\Delta t$ of the original system~\eqref{eq:sys}.

To prepare the dataset for training the neural network
map~\eqref{eq:netw_model}, we use the VAE encoder output $\mu(n)$ and
$\sigma(n)$ in two ways. First, each $u(n)$ is sampled from a random
distribution parameterized by $\mu(n)$ and $\sigma(n)$. Second, we set directly
$u_\mu(n)=\mu(n)$, shift it by one step, and compose pairs $(u(n),
u_\mu(n+1))$. Here $u(n)$ will be forwarded to the input of the trained network
and $u_\mu(n+1)$ will be its desired output. The full dataset includes $20$
repetitions of this procedure for the whole set of the embedding vectors $R(n)$,
so that there are $20$ slightly different sets of $u(n)$ each corresponding to
the same $u_\mu(n+1)$. The loss function is mean squared error
$(u'(n+1)-u_\mu(n+1))^2$, where $u'(n+1)$ is the output of the network for the
input $u(n)$.

These introduced fluctuations are necessary to compensate an unavoidable
imperfection of the training and provide robustness for the trained neural
network map. One can never train the network one hundred percent perfect, i.e.,
to make its outputs $u'(n+1)$ exactly be equal to the desired one
$u_\mu(n+1)$. When the neural network map~\eqref{eq:netw_model} is trained on
many trajectory cuts of a dynamical system with various parameter values, as it
is done in our previous works~\cite{NNDiscov22,NNNCPL2024}, the map is able to
reproduce well the dynamics even in spite of not perfect training. Now we have
only one trajectory corresponding to a single set of control parameter
values. Training the map~\eqref{eq:netw_model} without the random sampling of
$u(n)$ and setting $u(n)=\mu(n)$ instead is found to result in a fragile
reconstructed system. Its iterations are unable to follow the training
trajectory when even a small perturbation is added to the initial
point. However, the random sampling of $u(n)$ from the full Gaussian
distribution $N(\mu(n), \sigma(n))$ also results in inappropriate reconstructed
system that is unable to reproduce the modeled trajectory at all. The desired
neural network map that models the system~\eqref{eq:sys} well is obtained when
$u(n)$ are sampled from the truncated normal distribution $N(\mu(n), \sigma(n))$
with truncation at $0.1\sigma$. This particular value is not the single best
one. We checked truncations between $0.01\sigma$ and $\sigma$ and observed
similarly good results in a sense that the resulting maps reproduced the modeled
system.

The neural network map~\eqref{eq:netw_model} admits a vector $p$ that models
control parameters of the original dynamical system. In
papers~\cite{NNDiscov22,NNNCPL2024} we trained the map~\eqref{eq:netw_model} on
trajectories that were sampled at various parameter values of the modeled
system. These values where scaled to the interval $[-1,1]$ and accompanied the
corresponding trajectory samples as training data. In the present work the
source of the training data is a single trajectory of \eqref{eq:sys} so that
none of the parameters is emphasized over the others. For this reason we
consider $p$ as a scalar, i.e., $D_p=1$, and when the neural network map is
trained, the trajectory data are accompanied with a singe value $p=0.5$. This
constant is an arbitrary chosen value within the working range $[-1,1]$. The
seemingly more natural value $p=0$ cannot be chosen because then the
corresponding network weights would be zeroed out and no training would take
place.

When the training is finished we consider the dynamics of the resulting neural
network map~\eqref{eq:netw_model} at various $p$. It is interesting to establish
the correspondence between the regimes of the original system~\eqref{eq:sys} and
its reconstructed version~\eqref{eq:netw_model} at various parameter values.

\section{Dynamics of the neural network map}

We use the neural networks as described in Secs.~\ref{sec:VAE} and \ref{sec:ntw}
to reconstruct dynamics of the system~\eqref{eq:sys} from a single scalar time
series. We solve Eqs.~\eqref{eq:sys} numerically at fixed parameter set when
this system demonstrates bursting oscillations as in
Fig.~\ref{fig:sol_sherman_ode_o_burst}, then take a time series
$S(n)=S(t_0+n\Delta t)$ for one period of bursting, and construct
delay-coordinate embedding vectors $R(n)$ with the excessively large dimension
$D_e=32$. First these vectors are employed to train the VAE, and then the
state-space vectors $u(n)$ of the reduced dimension $D_u=3$ are reconstructed
from $R(n)$ by the encoder part of the VAE. In turn, the obtained $u(n)$ are
used to train the neural network map~\eqref{eq:netw_model}. During the training,
the trajectory data are accompanied by the parameter value $p=0.5$. In this
section we examine the regimes of the trained map~\eqref{eq:netw_model} at
various $p$ and contrast them with the regimes of the initial
system~\eqref{eq:sys}.

\begin{figure}
  \centering\includegraphics{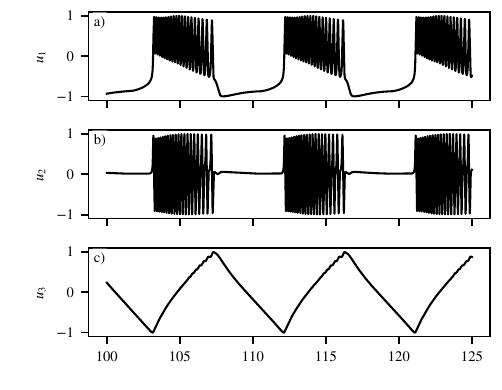}
  \caption{\label{fig:sol_sherman_ntw_o_burst}Iteration of the neural network
    map~\eqref{eq:netw_model} trained for the time series $S(t)$ in
    Fig.~\ref{fig:sol_sherman_vae_sv}(c). The parameter $p$ equals to the
    training value $p=0.5$. The starting point of the iterations is chosen
    arbitrarily.}
\end{figure}

\begin{figure}
  \centering\includegraphics{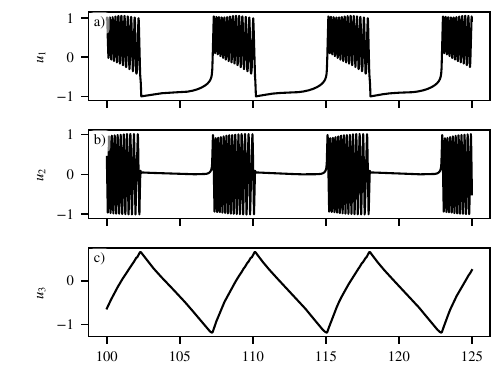}
  \caption{\label{fig:sol_sherman_ntw_o_burst_ii}Bursts produced as a result of
    iterations of the neural network
    map as in Fig.~\ref{fig:sol_sherman_ntw_o_burst} for $p$ beyond the training
    value, $p=0.8$.}
\end{figure}

\begin{figure}
  \centering\includegraphics{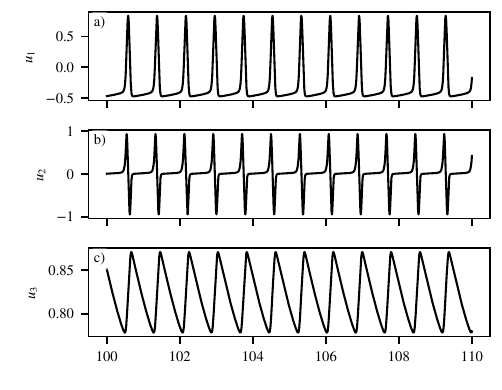}
  \caption{\label{fig:sol_sherman_ntw_o_spikes}Spiking in the neural network map
    as in Fig.~\ref{fig:sol_sherman_ntw_o_burst} at $p=0.2$.}
\end{figure}

Figures.~\ref{fig:sol_sherman_ntw_o_burst},
\ref{fig:sol_sherman_ntw_o_burst_ii}, and~\ref{fig:sol_sherman_ntw_o_spikes}
demonstrate dynamics of the neural network map~\eqref{eq:netw_model} trained for
the system~\eqref{eq:sys} at $V_S=-36$ and $k=0$. The initial trajectory is
shown in Fig.~\ref{fig:sol_sherman_vae_sv}(c). The data reconstructed by the VAE
encoder and are used for training the neural network map are shown in
Fig.~\ref{fig:sol_sherman_vae_sv}(d-f).

Figure~\ref{fig:sol_sherman_ntw_o_burst} is plotted at the point of training
$p=0.5$. Note that in this and in the following figures the starting point of
the iterations is chosen at random, not from the training data.  We see that the
trajectory generated by the neural network map reproduces the training data very
well, compare Fig.~\ref{fig:sol_sherman_ntw_o_burst} and
Fig.~\ref{fig:sol_sherman_vae_sv}(d-f). As previously discussed, the VAE is
unable to accurately reconstruct the precise structure of the variable $n$ of
the system~\eqref{eq:sys}, compare the original curve $n(t)$ in
Fig.~\ref{fig:sol_sherman_vae_sv}(b) and its reconstructed version in
Fig.~\ref{fig:sol_sherman_vae_sv}(e). The neural network map accurately recovers
this feature of the training data, compare
Fig.~\ref{fig:sol_sherman_ntw_o_burst}(b) and~\ref{fig:sol_sherman_vae_sv}(e).

It is of interest to observe the behavior of the map~\eqref{eq:netw_model} when
the value of $p$ is taken beyond the training
value. Figures~\ref{fig:sol_sherman_ntw_o_burst_ii} and
\ref{fig:sol_sherman_ntw_o_spikes} illustrate that varying $p$ produces a
comparable outcome to varying $V_S$ in the system~\eqref{eq:sys}. As illustrated
in Fig.~\ref{fig:sol_sherman_ntw_o_burst_ii} the system exhibits bursting at
$p=0.8$. Similarly, Fig.~\ref{fig:sol_sherman_ntw_o_spikes} reveals the presence
of spikes at $p=0.2$. These observations can be compared with
Figs.~\ref{fig:sol_sherman_ode_o_burst} and~\ref{fig:sol_sherman_ode_o_spikes},
which show bursting and spiking in the system~\eqref{eq:sys} at different values
of $V_S$. Also it is important to note that spikes were not demonstrated to the
neural network during training.

\begin{figure}
  \centering\includegraphics{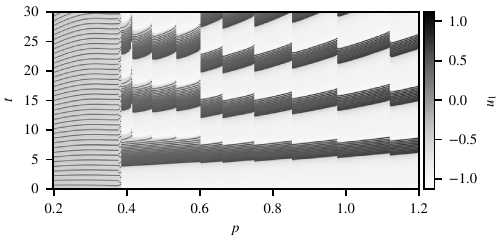}
  \caption{\label{fig:bif_diag_sherman_ntw_o}Diagrams of regimes of the neural
    network map~\eqref{eq:netw_model} trained for the system~\eqref{eq:sys} at
    $V_S=-36$ and $k=0$. The parameter $p$ is represented horizontally, with the
    vertical axis corresponding to time. Gray shades indicate values of the
    variable $u_1$, which reproduces the variable $V$ of the
    system~\eqref{eq:sys}. Compare it with the diagram in
    Fig.~\ref{fig:bif_diag_sherman_ode_o}.}
\end{figure}

Figure~\ref{fig:bif_diag_sherman_ntw_o} demonstrates the diagram of the regimes
of the neural network map~\eqref{eq:netw_model} trained for the
system~\eqref{eq:sys} at $V_S=-36$ and $k=0$. This figure is plotted in the same
manner as Fig.~\ref{fig:bif_diag_sherman_ode_o}. It is evident that the diagrams
for the system~\eqref{eq:sys} and its reconstructed model~\eqref{eq:netw_model}
exhibit a high degree of similarity. The most apparent and nonessential
distinction is the directionality of the parameter variations. An increase in
$V_S$ is associated with a decrease in $p$.

It was observed that varying $p$ is similar to varying $V_S$ in the modeled
system~\eqref{eq:sys}. However, this correspondence is not particularly
noteworthy, as bursts and spikes are the two most typical regimes of the system,
and varying other parameters will also result in oscillations of these two
types.

The system~\eqref{eq:sys} exhibits a single fixed point that within the
considered parameter range is always unstable at $k=0$. The modification at
$k=1$ results in the area where the fixed point becomes stable, thereby
initiating bistability between the bursting and the fixed point. In our previous
work~\cite{NNDiscov22} we demonstrated that the neural network
map~\eqref{eq:netw_model} can discover the fixed point. To train the map we used
various trajectory cuts of the system~\eqref{eq:sys} that never approached the
fixed point. The trained neural network map as expected was able to reproduce
the bursting and spiking dynamics, because this behavior were shown it in the
course of training. But also the map discovered the fixed point with its correct
stability properties, see Fig.~8 of the referred paper~\cite{NNDiscov22}. In the
present paper, we consider a more complex situation. The only information about
the system is one period of its bursting trajectory, which is displayed to the
neural network map via the single variable $S$. We train this network and aim to
ascertain which properties of the fixed point can be recovered by the network in
this case.

\begin{figure*}
  \centering\includegraphics{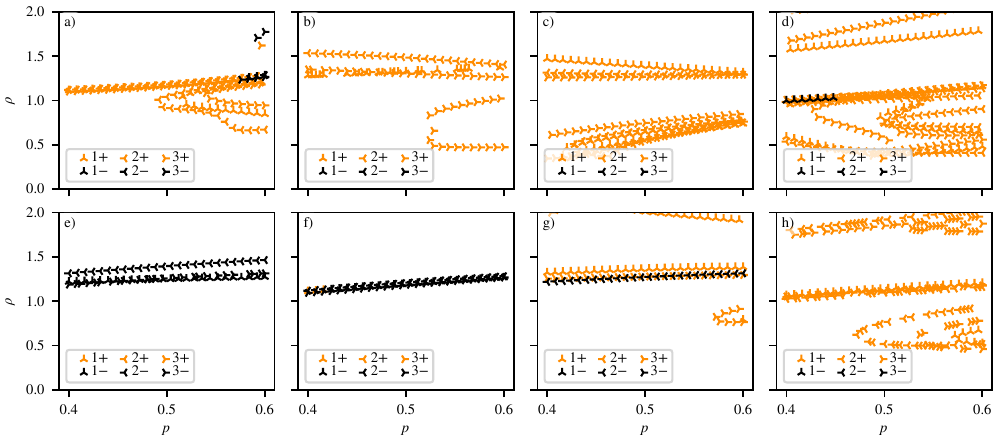}
  \caption{\label{fig:_ntw_fix_points}Distance $\rho$ to the origin of fixed
    points of neural network maps~\eqref{eq:netw_model} trained for different
    parameters $k$ and $V_S$ of the system~\eqref{eq:sys}. Along the horizontal
    axes are plotted the values of the parameter $p$. Panels (a--d) in the upper
    row correspond to the case $k=0$ and for panels (e--h) in the lower row
    $k=1$. Columns of panels from left to right demonstrate variation of $V_S$:
    (a) and (e) $V_S=-35.5$; (b) and (f) $V_S=-36$; (c) and (g) $V_S=-36.5$; (d)
    and (h) $V_S=-38$. Each panel contains three curves corresponding to three
    neural network maps trained for identical parameters and random initial
    network weighs. The curves are plotted with triangle markers $1$, $2$ and
    $3$. Black and orange colors of the markers indicate stability of the
    corresponding fixed point. Black color with ``$-$'' in the legends indicate
    stable fixed point, and orange marker with ``$+$'' depict unstable fixed
    points.}
\end{figure*}

Due to the fact that no fixed point information is explicitly presented to the
neural network during the training process, the resulting neural network map
usually has more than one fixed point. Some of these fixed points are located at
a considerable distance from the origin and thus they are not considered because
the bursting and spiking oscillations occur within the range $u_i\in [-1,1]$ as
a result of the preliminary rescaling of the training data. Moreover, the
distinct training runs may yield disparate sets of fixed points due to the
random initialization of the network weights. To identify nevertheless patterns
in the fixed points appearance for each parameter of the system~\eqref{eq:sys},
we will train three neural network maps that differ only in their initial random
weights. Thus, the parameter values of the system~\eqref{eq:sys} will be
considered in eight combinations: $V_S=-35.5$, $-36$, $-36.5$ and $-38$ at $k=0$
and $k=1$, for each parameter set the time series of $S(t)$ will be computed,
the VAE will be trained on it, and then three versions of the neural network map
will be trained for each dataset produced by the VAE encoder.

To illustrate the results, we plot the distance of the fixed point from the
origin, denoted by $\rho$, as a function of $p$ in the vicinity of the training
value $p=0.5$, see Fig.~\ref{fig:_ntw_fix_points}. Stable and unstable fixed
points are shown with black and orange colors, respectively, and additionally
marked by the symbols ``$-$'' and ``$+$'' in the legends. The three versions of
the neural network map are shown with markers of different shapes.

As one can see in Fig.~\ref{fig:bif_diag_sherman_ode_m}, the parameters
$V_S=-35.5$, $-36$, and $-36.5$ lay within the area of bistability when $k=1$,
while at $k=0$ the fixed point is always unstable,
Fig.~\ref{fig:bif_diag_sherman_ode_o}. The fixed points of the neural network
maps~\eqref{eq:netw_model} reconstructed for these parameter values at $k=0$ are
shown in Figs.~\ref{fig:_ntw_fix_points}(a, b, c). It is noteworthy that there
are no stable fixed points at $p=0.5$ and in the vicinity of this value, which
is consistent with the properties of the system~\eqref{eq:sys} at
$k=0$. Figures~\ref{fig:_ntw_fix_points}(e, f, g) corresponds to the maps
trained when the modeled system had a stable fixed point at $k=1$. One sees that
at $V_S=-35.5$ and $-36$, Figs.~\ref{fig:_ntw_fix_points}(e, f), respectively,
all three reconstructed maps also have a stable fixed point.  However, for the
maps trained at $V_S=-36.5$, Fig.~\ref{fig:_ntw_fix_points}(g), only one of them
exhibits the expected stable fixed point, while for two others the fixed point
is unstable even exactly at the training point $p=0.5$.

Regardless of $k$, the fixed point of the modeled system~\eqref{eq:sys} is
unstable at $V_S=-38$. Consequently, all reconstructed maps also lack stable
fixed points at $p=0.5$ and in the vicinity of this value, see
Figs.~\ref{fig:_ntw_fix_points}(d, h).

In conclusion, the training procedures under discussion do not consistently
recovers the stability property of the fixed point of the
system~\eqref{eq:sys}. However, a clear trend has been observed: in the
presented examples, only one of eight considered cases does not fully correspond
to the expected behavior: in Fig.~\ref{fig:_ntw_fix_points}(g) not all runs of
the training result in the expected stable fixed point.

\begin{figure}
  \centering
  \includegraphics{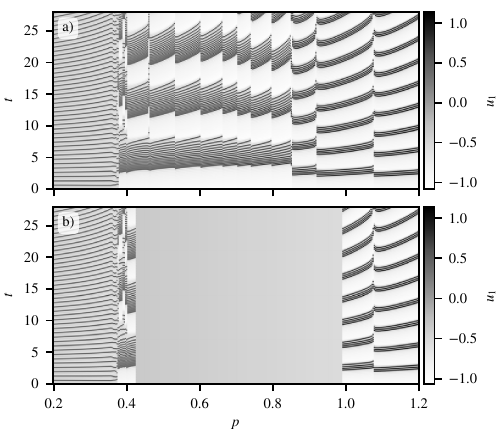}
  \caption{\label{fig:bif_diag_sherman_ntw_m}Diagram of regimes of the neural
    network map~\eqref{eq:netw_model} trained for the system~\eqref{eq:sys} at
    $V_S=-36$ and $k=1$. Panels (a) and (b) are computed as continuations of two
    solutions, the bursting and the fixed point, respectively, computed in the
    middle part of figure at $p=0.5$.}
\end{figure}

When the system~\eqref{eq:sys} is reconstructed at parameter values where its
fixed point is stable, say $k=1$, $V_S=-36$, and the trained map correctly
identifies the stability property of this point, a diagram of regimes with
bistability similar to that in Fig.~\ref{fig:bif_diag_sherman_ode_m} can be
plotted, see Fig.~\ref{fig:bif_diag_sherman_ntw_m}. In
Fig.~\ref{fig:bif_diag_sherman_ntw_m}(a) we compute a solution starting form an
arbitrary initial point at $p=0.5$ and then continue this solution to the left
and to the right. Then we find the stable fixed point at $p=0.5$ and again
continue it to the left and to the right,
Fig.~\ref{fig:bif_diag_sherman_ntw_m}(b). It can be observed that the fixed
point remains stable approximately for $p\in [0.4, 1]$, while outside of this
range, the oscillating solutions, burst and spikes, appear. It is notable that
this diagram is sufficiently similar to the diagram plotted directly for the
system~\ref{eq:sys}, see Fig. \ref{fig:bif_diag_sherman_ode_m}.

\section{Conclusion}

This work examines the reconstruction using neural networks of a family of
dynamical systems with neuromorphic behavior from a single scalar time
series. We consider a model of a physiological neuron built on the basis of the
Hodgkin-Huxley formalism. Taking its single scalar time series at fixed values
of the control parameters, we reconstruct a one-parameter family of dynamical
systems as a neural network map and study the correspondence of the behavior of
this family to the behavior of the Hodgkin-Huxley neuron for various parameter
values.

The general motivation of this work is a contribution to a theoretical framework
for processing experimental data. A typical situation when working with
experimental data is that there is a single record of an experiment from which
it is necessary to extract the maximum amount of information about the
full-scale system under the study. In this work, we simulate such a situation,
but for now we do not set a goal to reproduce it exactly, namely: our time
series is obtained as a result of solving a system of differential equations,
the data is recorded with a sufficiently small time step and is not noisy.

The reconstructed family of dynamical systems is a neural network, which after
the training operates as a recurrent map, i.e., a discrete-time dynamical
system, and has one control parameter. The reconstruction is performed for a
scalar time series of the original system in bursting mode.

It is shown that, despite the fact that the spiking oscillations are not
demonstrated to the network during training, the reconstructed neural network
map, when changing the parameter, demonstrates the transition from bursts to
spikes in a qualitatively similar way to how it occurs in the modeled
system. For the modeled system and the neural network map, diagrams or regimes
are represented that show the transition from bursts to spikes. The diagrams
correspond well to each other.

The situation is studied when a modification is introduced into the
Hodgkin-Huxley system, which makes it possible for the emergence of a
bistability regime of burst dynamics and a stable fixed point. The neural
network map was reconstructed from the bursting trajectory of such a system in
the bistability mode, and it was shown that it can also have a stable fixed
point. The represented diagrams of regimes for bistability demonstrate qualitative
agreement with similar diagrams of the simulated system.

The stability of fixed points of neural network map reconstructed for different
values of the parameters of the modeled system in the presence and absence of a
modification that generates bistability is studied. Note that during the
training process the fixed point was never demonstrated to the neural
network. It is shown that the nature of the stability of fixed points present in
the neural network maps trained in this way does not always strictly correspond
to the system being modeled. However, by performing the training procedure
several times for the same parameters of the modeled system, we showed that
there is a clearly visible tendency towards a much more frequent appearance of
fixed points with the ``correct'' character of stability.

When reconstructing a dynamical system from a scalar time series, the question
arises about the dimension of the reconstructed state-space vectors. In our
work, we construct for a time series delay-coordinate embedding vectors, while
setting their dimension to be excessively large, namely ten times larger than
the true dimension of the state-space vectors of the moderated system. Then
their dimension is reduced using a variational autoencoder, which in its latent
space restores the system state vectors. These vectors are subsequently used for
training the neural network map. The question of the correct choice of the
dimension of the reconstructed state-space vectors is discussed. It is shown
that the true dimension corresponding to the dimension of the modeled system can
be found on the basis of varying the dimension of the latent space and analyzing
the patterns in the appearance of new components, as well as by observing the
changes in the behavior of learning curves.

\begin{acknowledgments}
  This work was supported by the Russian Science Foundation, 20-71-10048,
  https://rscf.ru/en/project/20-71-10048/. The study of model dynamics was
  carried out within the framework of the project “Mirror Laboratories” HSE
  University (Section II).  
\end{acknowledgments}

\section*{Data Availability Statement}

The data that support the findings of this study are available within the
article.

\appendix

\section{\label{sec:sys_detail}Functions and parameter values of the
  Hodgkin-Huxley-type model}

The functions included to Eqs.~\eqref{eq:sys} are as follows:
\begin{subequations}
  \label{eq:origsys_funcs}
  \begin{gather}
    I_{Ca}(V)=g_{Ca} \, m_{\infty}(V) \, (V-V_{Ca}), \label{eq:origsys_ica}\\
    I_{K}(V,n)=g_{K}\, n\, (V-V_{K}), \label{eq:origsys_ik}\\
    I_{S}(V,S)=g_{S}\, S\, (V-V_{K}), \label{eq:origsys_is}\\
    I_{K2}(V)=g_{K2} \, p_{\infty}(V) \, (V-V_{K}), \label{eq:modif_cur}\\
    \omega_{\infty}(V) = \left( 1+\exp \frac{V_{\infty}-V}{\theta_{\omega}}
    \right)^{-1}\hspace{-1.2em},\hspace{1.2em} \omega=m,n,S, \label{eq:orig_gates}\\
    p_{\infty}(V) = \left( \exp \frac{V-V_{p}}{\theta_{p}}+ \exp
      \frac{V_{p}-V}{\theta_{p}} \right)^{-1}\hspace{-1.2em}. \label{eq:modif_prob}
  \end{gather}
\end{subequations}

Numerical values of the control parameters used in simulations of the
system~\eqref{eq:sys} can be found in Table~\ref{tab:param}.

\begin{table}[h]
  \caption{\label{tab:param}Numerical values of parameters of the model~\eqref{eq:sys}}
  \begin{center}
  \begin{tabular}{llll}
    \hline
    $\tau=0.02\,\text{s}$      & $\tau_S=35\,\text{s}$       & $\sigma=0.93$              & {} \\
    $g_{Ca}=3.6$               & $g_{K}=10$                  & $g_{S}=4$                  & $g_{K2}=0.12$ \\
    $V_{Ca}=25\,\text{mV}$     & $V_{K}=-75\,\text{mV}$      & {}                         & {} \\
    $\theta_{m}=12\,\text{mV}$ & $\theta_{n}=5.6\,\text{mV}$ & $\theta_{S}=10\,\text{mV}$ & $\theta_{p} = 1\,\text{mV}$ \\
    $V_{m}=-20\,\text{mV}$     & $V_n=-16\,\text{mV}$        & $V_{S}=-36\,\text{mV}$     & $V_{p}=-49.5\,\text{mV}$ \\
    \hline
  \end{tabular}
  \end{center}
\end{table}

\section{\label{sec:vae_detail}Structure of VAE in detail}

\begin{figure}
  \centering\includegraphics{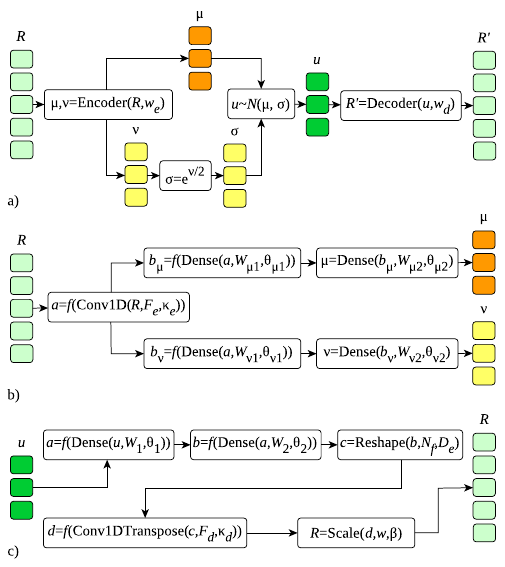}
  \caption{\label{fig:vae_arch}VAE: (a) general structure;
    (b) and (c) inner structure of the encoder and decoder, respectively. $R$ is
    the input vector; $\mu$ and $\sigma$ are parameters of the latent space; $u$
    is a vector of the latent space sampled from the Gaussian distribution
    $N(\mu, \sigma)$; $R'$ is the output vector; $w_e$ and $w_d$ are vectors of
    trainable parameters (neural networks weights) of the encoder and the
    decoder respectively:
    $w_e=\{F_e, \kappa_e, W_{\mu1,2}, \theta_{\mu1,2}, , W_{\nu1,2},
    \theta_{\nu1,2}\}$,
    $w_d=\{W_{1,2}, \theta_{1,2}, F_d, \kappa_d, w, \beta\}$.}
\end{figure}

The VAE we create for our problem is shown in
Fig.~\ref{fig:vae_arch}. Figure~\ref{fig:vae_arch}(a) demonstrates its common
structure and in Figs.~\ref{fig:vae_arch}(b, c) the inner structures of the
encoder and decoder are shown.

The input vector $R$ is forwarded to the encoder that maps it to a couple of
vectors $\mu,\nu\in\mathbb{R}^{D_u}$. Their dimension $D_u<D_e$ is the dimension
of the reconstructed dynamical system. The vector $\nu$ is treated as a
logarithm of a variance, and $\sigma=\euler^{\nu/2}$ is a standard
deviation. The vectors $\mu$ and $\sigma$ defines so called a latent
space. Elements of this space are vectors $u$ that represent the reduced form of
$R$. Vectors $u$ from the latent space are taken as a random samples of a normal
distribution $N(\mu,\sigma)$. So, when $\mu$ and $\sigma$ are generated for the
input $R$ the vector $u$ is sampled and then sent to the decoder. It maps this
vector to $R'$. The goal of VAE training is minimization of the discrepancy
between $R$ and $R'$ by tuning parameter vectors of the encoder $w_e$ and the
decoder $w_d$. Also a proper latent space structure has to be provided that is
achieved by simultaneous minimization of the Kullback–Leibler divergence between
the distribution parameterized by $\mu$ and $\sigma$ and standard normal
distribution~\cite{Kingma2019}. Thus the loss function for VAE training includes
two terms~\cite{Chollet2021}: the discrepancy between $R$ and $R'$ that we
compute as a mean squared error and the mean Kullback–Leibler divergence
computed via $\mu$ and $\nu$:
\begin{equation}
  \label{eq:vae_loss}
  L = \left\langle (R-R')^2 \right\rangle_{D_e} +
  K \left\langle \mu^2+\euler^\nu-\nu-1 \right\rangle_{D_u}
\end{equation}
The first and the second terms are averaged over $D_e$ and $D_u$ components of
the corresponding vectors, respectively. The coefficient $K$ is introduced to
balance minimization rates of the two terms: one sees that the first one can be
arbitrary small while the second one is at least $O(1)$. Thus the value of K has
to be of the order of the acceptable recovery error. We are going to continue
the training until the mean squared recovery error is of the order
$10^{-5}$. Thus we set $K=10^{-5}$.

The common architecture of VAE displayed in Fig.~\ref{fig:vae_arch}(a) is
typical for various topics, while the inner structures of the encoder and
decoder depend on the problem under consideration. In our case time series are
processed. There are three standard ways for dealing with them: one dimensional
convolutions, recurrent cells and transformers~\cite{Chollet2021}. Since we aim
to keep our neural networks as simple as possible, the convolutions are
employed. The encoder includes one convolution layer supplied with two dense
layers, Fig.~\ref{fig:vae_arch}(b). The decoder have to perform the encoder
operation in the reversed order. Thus it first includes two dense layers. Then
the flat output of these layers is reshaped to the matrix $N_f\times D_e$ to
fulfill the requirement for the input data of the transposed convolution
layer. Its output is one dimensional vector $d$ that is scaled as
$R = \layScale(d,w,\beta)=d w + \beta$ with scalar trainable coefficients $w$
and $\beta$ to fit the range of the desired output vector
$R$. Table~\ref{tab:encdec_dims} provides the description of the elements of the
encoder and decoder. See also the book~\cite{Chollet2021} for more descriptions
of the layers used.

\begin{table}
  \caption{\label{tab:encdec_dims}Structure of elements of the encoder and the
    decoder in Figs~\ref{fig:vae_arch}(b,c).  $N_k$ is convolution kernel size,
    $N_f$ is number of convolution filters, $N_{\mu 1}$ is size of the output of
    the first dense layer. $N_k=3$, $N_f=16$.  $N_{\mu 1}=512$,
    $f(\cdot)=\tanh(\cdot)$.}
  \begin{ruledtabular}
    \begin{tabular}{lll}
      Object & Type & Diemsnion \\
      \hline
      $R$ & row vector & $D_e$ \\
      $F_e$, $F_d$ & matrix & $N_k \times N_f$ \\
      $\kappa_e$ & row vector & $N_f$ \\
      $\kappa_d$ & scalar & {} \\
      $W_{\mu 1}$, $W_{\sigma 1}$ & matrix & $(D_e N_f)\times N_{\mu 1}$ \\
      $\theta_{\mu 1}$, $\theta_{\sigma 1}$ & row vector & $N_{\mu 1}$ \\
      $W_{\mu 2}$, $W_{\sigma 2}$ & matrix & $N_{\mu 1} \times D_u$ \\
      $\theta_{\mu 2}$, $\theta_{\sigma 2}$ & row vector & $D_u$ \\
      $W_1$ & matrix & $D_u\times N_{\mu 1}$ \\
      $\theta_1$ & row vector & $N_{\mu 1}$ \\
      $W_2$ & matrix & $N_{\mu 1}\times (D_e N_f)$ \\
      $\theta_2$ & row vector & $D_e N_f$ \\
      $w$, $\beta$ & scalars & {} \\
      \hline
      $f(\cdot)$ & \multicolumn{2}{l}{scalar function of
                    scalar argument} \\
    \end{tabular}
  \end{ruledtabular}
\end{table}

\section{\label{sec:netw_map_detail}Detail of the neural network map}

\begin{table}
  \caption{\label{tab:netw_dims}Structure of elements of
    Eq.~\eqref{eq:netw_model}.  For models in this paper $D_u=3$, $D_p=1$,
    $N_h=100$, $f(\cdot)=g(\cdot)=\tanh(\cdot)$, and $\chi=\Delta t=0.005$.}
  \begin{ruledtabular}
    \begin{tabular}{lll}
      Object & Type & Diemsnion \\
      \hline
      $u(n)$ & row vector & $D_u$ \\
      $u_i(n)$ & scalar & {} \\
      $\sansU{i}(n)$ & row vector & $D_u-1$ \\
      $p$ & row vector & $D_p$ \\
      $a_i$, $\mu_i$ and $\beta_i$ & row vectors & $N_h$ \\
      $b_i$ & column vector & $N_h$ \\
      $\gamma_i$ & scalar & {} \\
      $A_i$ & matrix & $(D_u-1)\times N_h$ \\
      $B_i$ & matrix & $D_p\times N_h$ \\
      \hline
      $f(\cdot)$ and $g(\cdot)$ & \multicolumn{2}{l}{scalar functions of
                                  scalar argument} \\
      $\chi$ & \multicolumn{2}{l}{small scalar constant}
    \end{tabular}
  \end{ruledtabular}
\end{table}

Structure of Eq.~\eqref{eq:netw_model} corresponds to a two-layer dense
network. The first layer obtains a scalar value $u_i$ and its output is a vector
of dimension $N_h$ that is the result of the expression
$f(u_i(n) a_i + \mu_i + g(\ldots))$. Layers like this are usually called
hidden. Its dimension $N_h$ determines a network information capacity and
performance. Unlike a generic two-layer network, the vector of biases $\mu_i$
get a correction form the output of an additional layer
$g(\sansU{i}(n) A_i + p B_i + \beta_i)$ that depends on other (non $i$th)
dynamical variables and control parameters. The second layer, i.e., network
output has a form $f(\ldots)b_i+\gamma_i$. It computes a scalar value that is
used as a correction to $u_i(n)$ to compute its value on step $n+1$:
$u_i(n+1)=u_i(n)+\chi(\ldots)$.

The neural network map~\eqref{eq:netw_model} is adopted for a required dynamics
during training procedure that consists of tuning the network weight parameters
\begin{equation}
  \label{eq:all_weights}
  w_m = \{
  a_i, \mu_i, A_i, B_i, \beta_i, b_i, \gamma_i\;|\; i=1,2,\ldots,D_u
  \}.
\end{equation}
The training vector $u(n)$ is forwarded to the network and the network computes
$u'(n+1)$. This vector is compared with the desired one $u(n+1)$. Their mean
squared error is a loss function of the training process. It is minimized in
course of training via gradient decent method~\cite{Chollet2021}. In actual
computations we used the modification of this method named
Adam~\cite{Kingma2014}. When training is finished and the network operates as a
discrete time dynamical system, we set $u'(n+1)=u(n+1)$.

Software implementation of the neural network map~\eqref{eq:netw_model} can be
done via the following operators. Operator $\layTake(u,i)$ extracts $i$th
elements of the vector $u$ and $\laySans(u,i)$ removes it and returns a vector
$\sansU{i}$ without this element. Operator $\layDense(x, W, b) = x W + b$
represents a fully connected (dense) layer without the activation: input row vector
$x$ is multiplied by a matrix $W$ and biased by a vector $b$. Finally,
square brackets $[\cdot,\cdot]$ denotes concatenation of two vectors or matrices.

\begin{figure}
  \centering\includegraphics{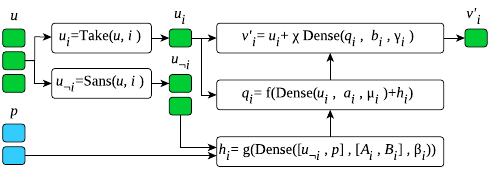}
  \caption{\label{fig:dynam_arch}Structure of the neural network for the $i$th
    variable of the map~\eqref{eq:netw_model}. Here $u\equiv u(n)$,
    $v'_i\equiv u'_i(n+1)$. The prime marks a resulting value computed by the
    network. This value in the course of training is compared with the
    ``correct''' value $u_i(n+1)$ from the training dataset.}
\end{figure}

In this notation neural network map~\eqref{eq:netw_model} for $i$th components
can be described as follows, see Fig.~\ref{fig:dynam_arch} and
Eqs.~\eqref{eq:netw_layers}. There are two input vectors, $u$ and $p$,
Eq.~\eqref{eq:netw_up}. Vector $u$ is split into a scalar $u_i$ and a vector
$\sansU{i}$ without the $i$th element, Eqs.~\eqref{eq:netw_split}. Layer in
Eq.~\eqref{eq:netw_lay1} takes $\sansU{i}$ and $p$ and compute a vector $h_i$ of
dimension $N_h$. This vector carries an information about the influence from the
control parameters $p$ and all elements of $u$ except the $i$th. This vector is
added to the output of the layer that processes the $i$th
component~\eqref{eq:netw_lay2}. At the output of this layer the vector $q_i$
appears that has the dimension $N_h$. It is passed to the last
layer in Eq.~\eqref{eq:netw_lay3} whose output is used to compute the solution at the
new time step $v'_i$. Doing these computations for $i=1,2,\ldots, D_u$ we obtain
the full vector at next time step $v'=u'(n+1)$.
\begin{subequations}
  \label{eq:netw_layers}
  \begin{gather}
    \label{eq:netw_up}
    p = \layInput(),\; u = \layInput(),\\
    \label{eq:netw_split}
    u_i = \layTake(u,i),\; \sansU{i} = \laySans(u, i),\\
    \label{eq:netw_lay1}
    h_i=g(\layDense([\sansU{i}, p], [A_i, B_i], \beta_i)), \\
    \label{eq:netw_lay2}
    q_i=f(\layDense(u_i,a_i, \mu_i) + h_i), \\
    \label{eq:netw_lay3}
    v'_i = u_i + \chi \layDense(q_i, b_i, \gamma_i).
  \end{gather}
\end{subequations}

As already mentioned above, the computed output $u'(n+1)$ is compared with the
desired one $u(n+1)$ during training, and when the training is finished we set
$u'(n+1)=u(n+1)$.

\bibliography{ann_dyn_recover}

\end{document}